\documentclass[preprints,article,accept,oneauthor]{mdpi} 
\pdfoutput=1
\input{frontdefs.tex}

\abstract{Biomedical research is intensive in processing information in the
  previously published papers. This motivated a lot of efforts to provide tools
  for text mining and information extraction from PDF documents over the past
  decade. The *nix (Unix/Linux) operating systems offer many tools for working
  with text files, however, very few such tools are available for processing
  the contents of PDF files. This paper reports our effort to develop shell
  script utilities for *nix systems with the core functionality focused on
  viewing and searching multiple PDF documents combining logical and regular
  expressions, and enabling more reliable text extraction from PDF documents
  with subsequent manipulation of the resulting blocks of text. Furthermore, a
  procedure for extracting the most frequently occurring multi-word phrases was
  devised and then demonstrated on several scientific papers in life sciences.
  Our experiments revealed that the procedure is surprisingly robust to
  deficiencies in text extraction and the actual scoring function used to rank
  the phrases in terms of their importance or relevance. The keyword relevance
  is strongly context dependent, the word stemming did not provide any
  recognizable advantage, and the stop-words should only be removed from the
  beginning and the end of phrases. In addition, the developed utilities were
  used to convert the list of acronyms and the index from a PDF e-book into a
  large list of biochemical terms which can be exploited in other text mining
  tasks. All shell scripts and data files are available in a public repository
  named \pp\ on the Github. The key lesson learned in this work is that
  semi-automated methods combining the power of algorithms with the
  capabilities of research experience are the most promising for improving the
  research efficiency.}


\begin{document}

\section{Introduction}

There is a growing interest to automate the consumption of scientific knowledge
to accelerate and automate research discoveries \cite{kitano2016, king2018,
  loskot2018}. Semantic enrichment and effective representation models of
research objects, their automated discovery and reuse can facilitate more
effective collaboration between the humans and machines \cite{gomez2017}.
Scientific papers are used the primary means for storing and sharing the
research findings and knowledge. The vast majority of scientific papers are
available in portable document format (PDF). This format was developed in the
early 90's by Adobe to efficiently represent information contents on a page for
the archival and presentation purposes. Unfortunately, the PDF does not support
information extraction and subsequent information processing. For general
scientific papers, the document elements of interest are metadata such as the
paper title, authors and their affiliations, abstract, keywords, section titles
and the corresponding full texts, figures, tables and their captions, and the
list of references. For papers in chemistry, biology and medicine, the key
elements also include names of chemical compounds, diseases, proteins, species,
and genes. In engineering, mathematics and physics, mathematical expressions
are often crucial for understanding the papers \cite{iwatsuki2017,
  udrescu2020}.

In this paper, we introduce several shell-script utilities newly developed for
processing text contents of PDF documents. These utilities were bundled as \pp\
in order to emphasize that they are aimed at processing the text contents in
scientific papers. The shell scripts are commandline programs to be run in a
terminal on a Unix or Linux operating system (OS). The implementation strategy
was inspired by the popular
\x{pdfjam}\footnote{\url{https://github.com/rrthomas/pdfjam}} program which is
a widely used shell script for manipulating pages of PDF files. The \x{pdfjam}
provides a simplified interface to the \LaTeX package
\x{pdfpages}\footnote{\url{https://www.ctan.org/tex-archive/macros/latex/contrib/pdfpages}}.
More importantly, \x{pdfjam} has become available in the repositories of many
common Linux distributions including Ubuntu, Debian, CentOS and Fedora.

The current development stage of \pp\ did not reach the maturity required to be
accepted by the Linux repositories. The initial source codes of \pp\ was
released on the Github\footnote{\url{https://github.com/ploskot/pdfPapers}}
under the GNU/GPL 2
license\footnote{\url{https://www.gnu.org/licenses/old-licenses/gpl-2.0.en.html}}
to support its future open source development. The main objective of \pp\
utilities is to improve the processing workflow of information extraction from
PDF files. The \pp\ improves and adds new functionality to \x{pdftotext} which
is likely the most reliable and the most commonly used tool for text extraction
from PDF documents available on the Linux OS. The \x{pdftotext} program is
distributed as one of the core utilities in the
\x{poppler}\footnote{\url{https://pypi.org/project/poppler-utils}} library
developed for PDF file conversion, manipulation and rendering. More
specifically, the main functionality added to \x{pdftotext} by \pp\ is handling
special characters and non-typical encodings, joining words and sentences split
across lines, columns, blocks and pages, searching the extracted text using
case-insensitive regular expressions combined with logical operators, and
generating the term-frequency (TF) statistics of multi-word phrases. The
extracted text is partitioned into logical units referred to as blocks. The
text blocks are defined by the internal structure of the PDF file, i.e., the
blocks correspond to the layout of page elements defining the page content. The
page layout is determined by the program which created the PDF file. For
instance, a single paragraph of text may be spread over several blocks. The
\pp\ assigns each text block with a unique identifier, so the blocks can be
copied, moved, deleted, concatenated, sorted, filtered, and eventually merged
into a label-free text file as the input for subsequent text mining algorithms.
For many PDF files, it is often the case that only a small number of blocks on
every page is relevant while all other blocks can be discarded. The \pp\ can
visualize the block layout on every page to aid the decision which text blocks
should be kept.

The text file produced by \pp\ is more reliable for subsequent text mining than
the raw text output produced by the standard \x{pdftotext} utility. The
multi-word phrase identification problem assumed in this paper is one of many
applications enabled by the reliable PDF to text conversion. Our experiments
suggest that the TF analysis of key phrases is sufficient to provide good
understanding of the contents and of the focus of the scientific paper even
without considering a corpora of papers and with no regard to the prior domain
knowledge represented as the controlled vocabulary, the list of domain terms or
otherwise. This approach facilitates more efficient reading of scientific
papers by individual researchers on their personal computers. Furthermore, our
findings indicate that key phrases in scientific papers are context dependent,
the stop-words should be removed after and not before the search for key
phrases, and that the relevant multi-word phrases can be reliably detected by
adding words to the previously found shorter phrases.

The rest of this paper is organized as follows. Section 2 surveys the existing
text mining approaches and tools for keyword and knowledge extraction from
biomedical papers. Section 3 outlines our methodology for identifying important
multi-word phrases in scientific papers. A necessary technical background to
understand challenges of text extraction from PDF files is also given, and the
\pp\ utilities and their implementation are summarized. The results of
identifying key phrases of up to 4 words in 5 selected biological papers, and
an example of creating a list of biological terms from the e-book list of
acronyms and the index are presented in Section 4. Our findings are discussed
and evaluated in Section 5. The paper is concluded in Section 6.

\section{Text Mining of Biomedical Documents}

Keywords can direct researchers to important parts of the document. Keywords
can be also utilized to produce document summary, perform topic classification,
name entity recognition and other such tasks. The concept of keywords is
intuitive, but it is difficult to define objectively \cite{rodriguez2009}. The
strategies how authors assign keywords to their papers is investigated in
\cite{babaii2013}. It has been found that the keywords selection is strongly
biased by the authors' background and expertise.

A recent very comprehensive survey of the keywords extraction methods and
issues appeared in \cite{firoozeh2019}. The survey attempts to define a
`keyness' of keyword, and how it can be related to different text features. The
keywords or keyphrases are assumed to be lexical units which can best represent
the document. The keyword selection can be made more objective by considering
the exhaustivity, specificity, minimality, impartiality, representativity
well-formedness, citationess, conformity, homogeneity and univocity of
keywords.

It has been recognized early on that important keywords in scientific papers
reflect their frequency of occurrence with respect to a domain-specific keyword
distribution \cite{andrade1998}. The domain distribution allows calculating the
likelihood score for every word in the text. The fact that keywords are
appearing statistically more often can overplay their differences rather than
account for their lexical similarities and while neglecting their semantic
differences \cite{baker2004}. In order to avoid over-interpretation or
under-interpretation of keywords, it is recommended to study dispersion
patterns, concordances and clusters of keywords, and to utilize annotated
texts. The TF distribution across a corpus of documents for keywords
identification is studied in \cite{kang2005}. Statistical approaches for
keywords identification based on their frequency of occurrence are reviewed in
\cite{kilgarriff2009}.

The domain-independent scoring of words using so-called C/NC-values aims at
enhancing the simple frequency of occurrence based identification
\cite{frantzi2000}. The semantic similarity is combined with the word frequency
in \cite{haggag2013}, and with a complete lexical database of English language
in \cite{haggag2014} to identify the document keywords. These approaches,
however, require semantic labeling of words which significantly complicates the
implementation.

The features which can be exploited for keyword extraction are enumerated in
\cite{kaur2010} including the frequency of occurrence, identification of nouns,
the presence of upper-case letters, the use of specific font shapes and faces,
the length of sentences they appear in, the presence of cue-words within the
same sentence, a relative position of the keyword and its sentence within the
paragraph, and the features based on conditional random-fields. The sets of
predefined keywords can be used to calculate the importance weights of other
words \cite{liang2017}. There have been also efforts to patent keyword
extraction methods \cite{patent1}.

The multi-word phrases are distributed differently in different sections of
scientific papers \cite{shah2003}. For instance, it was found that some key
phrases may not be present in the abstract. Multi-word phrases within the text
to extend the standard bag-of-words (BoW) approach are identified in
\cite{wang2018} by generative probabilistic models. The multi-word phrases are
then used to construct knowledge graphs representing the document. In
\cite{firoozeh2019}, it is reported that single token keywords usually account
for $17-20\%$, two-token keywords for $53-61\%$ and three-token keywords for
$21-18\%$ of all keywords. However, other references report that the key
phrases of more than 3 tokens can represent as many as $50\%$.

The emergence of open access publications has enabled more reliable keyword
identification using full-text articles than assuming only the abstracts
\cite{lin2009, cohen2010}. This has been confirmed by mining over 16 million
full-text biomedical papers and automatically extracting protein-protein,
disease-gene, and protein subcellular associations as the named entities in the
papers \cite{westergaard2018}. Reference \cite{dai2010} suggests to identify
useful terms by comparing the terms mentioned in abstract with their
occurrences in the rest of the paper. Using the Medical Subject
Headings\footnote{\url{https://www.nlm.nih.gov/mesh/index.html}} (MeSH)
thesaurus of biomedical terms and their frequency of occurrence, it was
reported in \cite{schuemie2004} that the keyword density is the greatest in
abstract followed by the results section whilst each section contains $30-40\%$
of information unique to that section.

The automated extraction of topic keywords of biomedical documents and their
classification according to MeSH is considered in \cite{garten2010}. The
automated classification of research articles assuming their abstracts on
\x{PubMed}\footnote{\url{https://pubmed.ncbi.nlm.nih.gov}} is performed in
\cite{simon2019}. The Jensen-Shannon divergence and cosine similarity are used
to cluster keywords in \cite{wartena2008}. Their performance is evaluated on
categories of Wikipedia articles. The similarity of words can be also measured
by the Jaccard coefficient as proposed in \cite{niwattanakul2013} to evaluate
the document keywords against the index terms. It is shown in
\cite{trieschnigg2009} that retrieval of relevant documents is improved by
assigning MeSH terms also to the information queries. Different systems for
automated assignment of MeSH terms to scientific texts were compared with the
manual assignment in \cite{trieschnigg2009}. The keyword matching between the
query and the documents has been described in \cite{rebholz2012} to aid
biomedical research via integrative biology. The assignment of MeSH terms to
biomedical articles is conceived as a ranking problem in \cite{huang2011}.

More generally, text mining methods for system biology enable going beyond
simple word searches \cite{ananiadou2006, hassani2020}. The main tasks of
biomedical text mining are reviewed in \cite{rodriguez2009}. The surveys of
text mining strategies for information extraction from scientific literature
can be found in \cite{nasar2018} and \cite{salloum2018}. Reference
\cite{krallinger2017} provides a comprehensive review of text mining methods
for chemistry. Complete text mining workflows for cancer system biology are
reviewed in \cite{zhu2013}. Combining text mining with annotated experimental
data for hypothesis generation and biological discovery is considered in
\cite{jensen2006} and \cite{natarajan2006}. Text mining for discovery of
biological interactions and hypothesis generation is considered in
\cite{li2013}. Adverse drug reactions are inferred from the literature using
the text mining methods in \cite{shang2014}.

A corpus of 97 fully annotated biomedical articles is announced in
\cite{verspoor2012} to serve as a benchmark for evaluating the performance of
different text mining tools as demonstrated for sentence splitting,
tokenization, syntactic parsing, and named entity recognition applications.
More importantly, it was found that the performance of trainable machine
learning methods may differ greatly if used on different data sets. Deep
learning for text feature extraction including keyword identification has been
considered in \cite{liang2017} and for the named entity recognition in
\cite{habibi2017, dang2018, lee2020}. A tool for chemical entity recognition in
texts is presented in \cite{rocktaschel2012}.

Automated classification of sentences into 11 core scientific concepts is
performed in \cite{liakata2012} using support vector machines and conditional
random fields. It has been found that the most discriminatory features for this
type of classification are grammatical dependencies between single word and
two-word keywords. The conditional random-fields are used in
\cite{hirohata2008} to perform the context-dependent classification of
sentences in abstracts of scientific papers. The same problem was addressed in
\cite{ruch2007} using trained Bayesian classifiers.

The training data independent categorization of biomedical texts according to
MeSH terms and the Gene Ontology is presented in \cite{ruch2006}. The
identification and subsequent classification of 10 distinct argumentative
schemes typically used in genetic research papers have been implemented in
\cite{green2015}. The 4 types of binary argumentative relations between
sentences were identified in \cite{kirschner2015} to obtain the graph
structures of argument in scientific papers.

Automatic term recognition, document clustering, classification and
summarization can improve the efficiency of performing systematic reviews of
scientific literature \cite{thomas2011, jonnalagadda2015, nie2017, loskot2019}.
A semi-automated screening of biomedical literature to identify relevant papers
has been implemented in \cite{wallace2010} using supervised machine learning.
Reference \cite{nie2017} presents a text mining framework for bibliometric data
to identify research trends and to design research.

There are online tools to define properly formed keywords for scientific
papers\footnote{\url{http://ulib.iupui.edu/keywords}}, and to obtain the word
lists of biomedical keywords and terms such as diseases, proteins, genes,
chemicals, cell lines and
species\footnote{\url{https://corposaurus.github.io/corpora}}. The online
DeCS/MeSH service\footnote{\url{https://decs.bvsalud.org/en}} allows to search
the structured MeSH descriptors to index biomedical articles. The command-line
keyword generator\footnote{\url{https://lab.kb.nl/tool/keyword-generator}}
finds the most likely single-word keywords in a corpus of documents using
either the TF or unsupervised latent Dirichlet allocation (LDA) machine
learning model.

The \x{PDFX} online utility extracts logical units from a PDF file by first
building a geometric model of every page containing textual and bitmap
elements. The elements are merged into logical units using their location
information on the page as well as using the font properties
\cite{constantin2013}. An open source layout-aware text extraction
utility\footnote{\url{https://github.com/BMKEG/lapdftext}} from PDF files was
reported in \cite{ramakrishnan2012}. The extraction is performed in blocks. The
blocks are then classified into logical units, and reordered to create an
appropriate reading flow. There are many other text extraction utilities from
PDF files such as
\x{TextFromPDF}\footnote{\url{https://github.com/mihailsalari/TextFromPDF}},
\x{pdftxt}\footnote{\url{https://pypi.org/project/pdftxt}},
\x{pdflines}\footnote{\url{https://github.com/proger/pdflines}}, and
\x{PDFBox}\footnote{\url{https://pdfbox.apache.org}}.

\x{TerMine}\footnote{\url{http://www.nactem.ac.uk/software/termine}} is an
online service for multi-word keywords identification. It can also utilize the
dictionary of acronyms.
\x{BioText}\footnote{\url{http://biosearch.berkeley.edu}} is a web-based
application for searching abstracts, figure captions as well as full texts in
over 300 open access biological journals \cite{hearst2007}.
\x{BioReader}\footnote{\url{https://services.healthtech.dtu.dk/service.php?BioReader-1.2}}
is a web-based utility for automatically searching and classifying papers based
on their abstract in \x{PubMed} database.

\x{Textpresso}\footnote{\url{https://textpressocentral.org/tpc}} is an online
tool for full-text annotations via keyword queries and semantic categories. The
\x{SAPIENT}\footnote{\url{http://www.sapientaproject.com/software}} software is
a tool for automated annotations of sentences assuming the defined core
scientific concepts.
\x{Tagcorpus}\footnote{\url{https://github.com/larsjuhljensen/tagger}} is a C++
program to find the named entities of proteins, species, diseases, tissues,
chemicals and drugs in a corpus of documents.
\x{LINNAEUS}\footnote{\url{http://linnaeus.sourceforge.net}} is a dictionary
based utility for the name recognition of biological species.

\textsc{sciBert}\footnote{\url{https://github.com/allenai/scibert}} is a deep
learning model trained on a large corpus of scientific papers which can be used
for sentences annotations and classifications.
\x{CERMINE}\footnote{\url{http://cermine.ceon.pl}} is a machine learning based
system for automated extraction of metadata from scientific papers including
authors names and affiliations, journal name, journal volume and number, and
the list of references. In this regard, this utility appears to enhance the
capabilities of \x{PDFX}.

An open source software for comprehensive text analysis referred to as General
Architecture for Text Engineering\footnote{\url{https://gate.ac.uk}} (GATE)
has been under the development for nearly past 20 years. The usability of GATE
was demonstrated in \cite{cunningham2013} for genomic-wide cancer mutation
associations, medical records analysis, and for drug-related searches. It is
concluded that text mining for life sciences and medical applications can be
made to be well-defined and reproducible.

\section{Methodology}

Before describing our strategy for identifying multi-word phrases, and its
implementation as a collection of shell script utilities, the main challenges
of extracting text from PDF documents are reviewed. The extraction of text from
PDF is a necessary step to enable processing the information contents of
scientific papers.

\subsection{Text extraction from PDF}

The PDF file format has been developed in the 90's by Adobe to describe page
contents which can be flexibly and precisely rendered at appropriate resolution
and scales on a variety of media. The PDF had replaced then prevailing
postscript page description language. However, unlike postscript, PDF is
missing many general features of programming languages as it focuses on its
single main purpose, i.e., efficiently describing the page content. Moreover,
unlike postscript, PDF files can directly render the selected page without
requiring to rebuild the contents of all the preceding pages. The page contents
are stored in a dictionary normally located at the end of the PDF file. The
dictionary can be optimized, e.g., linearized and compressed for a better
efficiency. The PDF documents can embed interactive forms, multimedia as well
as fonts for the characters used in the document.

The page presentation focus of PDF is very suitable for archiving purposes and
for consuming their information content by human readers. However, in the era
of automated information processing, the PDF format is much less suitable. The
content elements in a PDF document can be placed on pages in an arbitrary order
with no regard to a logical structure of the document or the natural reading
flow. For instances, a single paragraph of text may consist of multiple parts
which are rendered in any order. Since the logical structure of a document is
not available in the PDF file, it must be inferred. For example, the paragraphs
and other text units can be inferred from the elements locations on the page,
the inter-character spacing, and other font properties.

Another challenge in extracting text from PDF files is the use of special
characters and different character encoding schemes. The special characters
from different languages can be transliterated, or completely removed if they
are isolated, e.g., used as mathematical symbols. However, converting the
document characters from one encoding into another can be sometime problematic.
It is recommended to use UTF-8 (Unicode Transformation Format 8-bit) encoding,
since it can efficiently represent over 1.1 million of valid characters using 1
to 4 bytes (8-bit values).

The next challenge is joining the words which are split across lines or even
pages. This is usually straightforward if the word is split across consecutive
lines using a dash delimiter. However, the word can become permanently split if
the delimiter has been replaced with a space during text extraction or
character encoding changes. Such cases are very difficult to detect and
rectify. Furthermore, it is often desirable to extract full sentences even if
they are split across multiple lines or even pages. The beginnings and ends of
sentences are usually detected by a set of delimiting characters such as dot,
and exclamation and question marks which are preceded by a lower-case letter
and followed by a space and an upper-case letter. Other separators such as
comma, colon and semicolon can usually be removed unless they are utilized in
semantic analysis of sentences. Our experiments suggest that it is best to
remove all end-of-line characters within the paragraphs before joining the
split words and sentences. It is also desirable to replace all repeated
whitespace characters including spaces and tabs with a single space.

\subsection{Identification of multi-word phrases}

Our objective is to detect relevant or important multi-word phrases within the
text extracted from a single PDF document. These phrases are deliberately not
referred to as the most relevant or the most important, since the phrase
relevance and importance is strongly dependent on the context and the task
we are trying to accomplish. For instance, the same scientific paper may be
added to one survey covering a certain topic based on one set of keywords, and
then to another survey on a different topic assuming a different set of
keywords. Both these sets of keywords which may best describe the paper can
have little or even no overlap. However, if one set of keywords is the subset
of another set of keywords, it is sensible to demand that the larger set is a
better description of the paper than the smaller set. In this case, it may be
possible to add more keywords to the set until they become a sufficiently good
representation of the paper. This argument also implies that keywords can be
assigned scores, so they can be sorted in terms of their relevance or
importance.

An expert may assign synonymous terms to the paper that do not appear directly
anywhere in the text. For example, these terms may be more appropriate
terminology normally used within a given domain. Furthermore, similarly to
uncertainty in determining how many keywords should be used to represent the
paper, there is uncertainty in how many neighboring words should be assumed in
identifying the relevant multi-word phrases. It is clear that a whole sentence
is more accurate description of the paper than its part, provided that the
efficiency of description may be ignored. However, unlike the problem of
determining the sufficient number of keywords which likely depends on the paper
length, its structure and its information content, the multi-word phrases of
interest are likely to consist of only a small number words. Whether a shorter
phrase is more relevant to better describe the paper than a longer phrase is
again context dependent. Our experiments indicate that shorter phrases are
preferred if more general scope of the paper is of interest whereas longer
multi-word phrases tend to create a more narrow description of the paper.

In this paper, the keyword identification issues outlined above are addressed
pragmatically by assuming the frequency of occurrence, i.e., the TF of
individual keywords as well as of multi-word phrases as the main metric to
enable their ranking. Since keywords with the largest TF are usually contained
in the paper title, scoring the keywords by their TF can become easily biased
due to many titles appearing in the list of references. Although it may be
possible to detect and exclude the list of references when calculating the TF
scores, we have investigated the strategy of calculating the spread of
candidate keywords throughout the whole paper. In particular, if $s_i$ denotes
the number of words between the keywords at word locations $l_i$ and $l_{i+1}$
in the paper, respectively, the spread of such keyword with its $N$ occurrences
in the paper normalized by the mean value is computed as,
\begin{equation}\label{eq:1}
  S = \frac{1}{N-2} \sum_{i=1}^{N-1} \left(s_i - \frac{1}{N-1}
    \sum_{i=1}^{N-1} s_i \right)^2 \ \left( \frac{1}{N-1} \sum_{i=1}^{N-1}
    s_i \right)^{-2}
\end{equation}
where $s_i=l_{i+1}-l_i-1$. The smaller the normalized spread $S$, the more
evenly the keyword is distributed throughout the paper, and the more likely
such keyword is sufficiently important, so it is mentioned in different parts
of the paper. However, we did not observe major changes in the ordered lists of
identified keywords by tweaking the scoring metric, although local changes in
the list do appear if the scores are adjusted.

The word stemming and word normalizations have not been considered in our
implementation, since they do not fundamentally affect our keyword search
strategy. On the other hand, the case-insensitive search is assumed. It can be
implemented by either converting all letters in the extracted text to
lower-case, or by setting the case-insensitive options in calling the shell
script commands. The stop-words should be removed, however, only under defined
circumstances. The rules adopted for identifying multi-word keywords which were
implemented in our shell scripts can be summarized as follows. It is assumed
that the full-text extracted from a PDF file was already curated for special
characters, split words and split sentences.

\begin{enumerate}
\item The objective is to identify the multi-word phrases having the largest
  frequency of occurrence.
\item The phrases are searched hierarchically at multiple levels. Starting at
  level 1, the one-word keywords are obtained, then at level 2, the two-word
  keywords are identified and so on.
\item The candidate keywords in the next level can be enumerated by appending or
  prepending single words to the phrases in the current level.
\item The number of phrases considered at a given level should be larger than
  at the previous level.
\item The phrases can contain stop-words provided that the stop-words are
  neither their first nor the last word. However, the phrases can contain
  stop-words anywhere, provided that these phrases are used to generate new
  extended candidate phrases at the next level, and not used as the phrases
  identified at the current level.
\item It is desirable to manually prune the phrases generated at every level in
  order to prevent the unlikely phrases to propagate to the next level.
\end{enumerate}

The procedure for identifying multi-word keywords consists of the following
steps. The most frequently occurring single words are identified in the first
level while all stop-words are excluded. A certain desired number of these
words can be declared as the single-word keywords. However, many more single
words from level 1 should be considered in level 2 to generate the candidate
two-word phrases by prepending and appending one neighbouring word from the
text to each of the single-word keyword. The two-word phrases having a
stop-word as the first or the second word can be excluded. A certain desired
number of the most frequently occurring two-word phrases can be assumed to be
the most relevant in level 2. Many more two-word phrases should be assumed in
level 3 to prepend and append single neighbouring words from the text to find
the most relevant three-word phrases in level 3. These steps are repeated to
find four-word phrases in level 4 and so on. The procedure can be usually
terminated after generating four or five-word phrases. The resulting output are
several sets of phrases consisting of 1 to 4 or 5 words having the greatest
frequency of occurrence in the text, and which do not contain stop-words as
their first or last word.

\begin{figure}[H]
  \centering
  \includegraphics[scale=1.35]{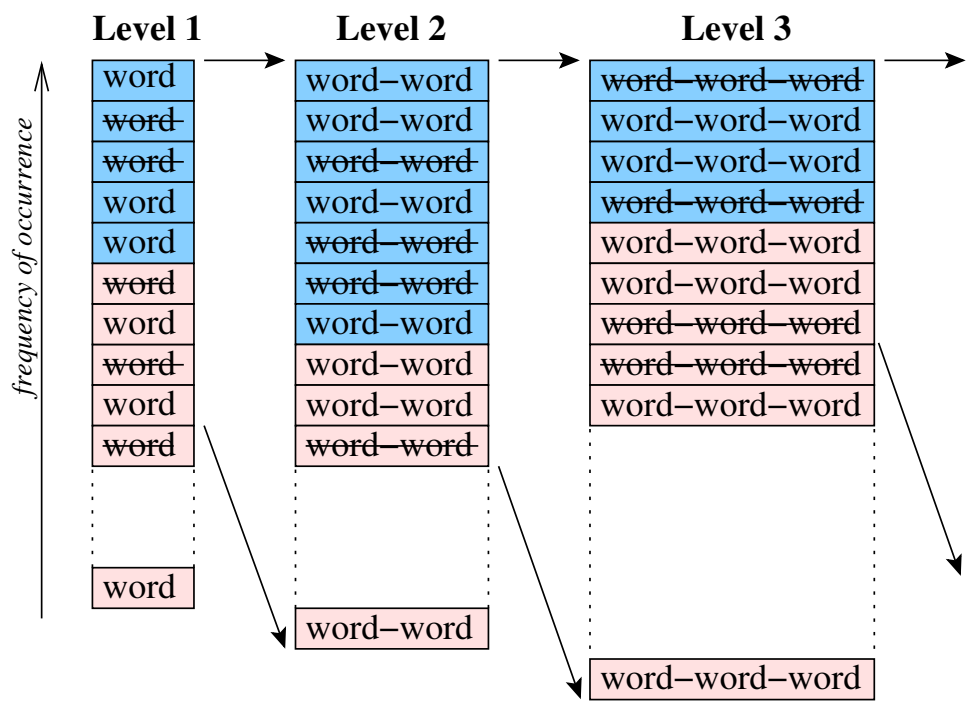}
  \caption{The proposed hierarchical iterative procedure for generating the
    relevant multi-word phrases.}\label{pict2a}
\end{figure}

The selection of important multi-word phrases at each level is depicted in
\fref{pict2a}. At each level, the phrases consisting of one or more words are
ranked by their frequency of occurrence. The blue cells represent the phrases
which are selected at each level. However, the search for phrases at the next
level requires that many other phrases in pink cells are considered too. The
ratio of the number of blue cells to the number of pink cells can be $1:10$ or
even smaller. Although the overall number of candidate phrases (blue and pink
cells) growth rapidly at each next level, the number of meaningful phrases in
blue cells is quickly reduced after level 2. The crossed phrases (words) can be
excluded automatically (e.g., they do not satisfy the stop-word constraint), or
they can be excluded manually by inspection (e.g., they may be outside the
intended scope or context).

Any scoring system to sort the candidate keywords is an attempt to estimate the
likelihood that the considered keywords are relevant or important in a given
context and a given application. Provided that the number of phrases considered
from the previous level is sufficiently large, the actual choice of the scoring
metric appears to be less important. The number of candidate phrases grows
significantly at each level. On the other hand, the number of meaningful
phrases having some minimum number of occurrence decreases rapidly at each
level. These two opposing phenomena usually yields the maximum number of
meaningful keywords for two-word phrases.

Some frequent words can have a common prefix. There are cases where it makes
sense to merge such words. However, this affects the generation of longer
phrases using the proposed procedure, so the word stemming has not been
considered. Nevertheless, manually pruning the generated lists of phrases
proved to be a very robust strategy to obtain the satisfactory results. For
instance, the two-word phrase, `in vivo', would normally be discarded, since
the first word is a stop-word, however, manual pruning can keep this term in
the list of candidate two-word phrases.

\subsection{Implementation as shell-scripts}

Our implementation was inspired by \x{pdfjam}. It is a shell script for
manipulating pages of PDF files which is available in most Linux distributions.
In general, the Linux shell has been designed from the very beginning to be
strongly oriented on text processing. There are many standard tools available
in every Linux shell that are specialized for such processing. The most
commonly used are these utilities:
\begin{enumerate}
\item \x{tr}: a utility to translate and delete characters in text files
\item \x{grep}: a utility to select lines in a text file that match given
  pattern
\item \x{sed}: a streaming editor for filtering and transforming text streams
\item \x{awk}: a text processor implementing a full programming language for
  patterns matching and text processing.
\end{enumerate}
It should be noted that text processing in the Linux shell is line oriented,
i.e., a text file is processed line by line. This may create problems when the
textual information to be processed is spread over multiple lines, and e.g.
paragraph by paragraph processing instead of the default line by line
processing is required. There are strategies for implementing multiline
processing of text, and it has been done in our scripts, but it makes the
scripts more complicated.

The shell scripts reported in this paper were developed and tested in BASH
(Bourne Again Shell) version 5.0.17 on Fedora Linux Workstation version 33.
These scripts are developed and distributed under the name \pp. In addition to
the above mentioned standard Linux shell programs, our implementation utilizes
the following shell script programs which may not be installed by default:
\begin{enumerate}
\item \x{poppler-utils}: a collection of Python utilities for manipulating and
  converting PDF files which are based on the open-source Poppler PDF library
\item \x{convert}: a powerful image converter which can transform many
  different file formats; it is included in \x{ImageMagick} collection of tools
\item \x{gnuplot}: an interactive plotting program supporting many different
  output devices and formats
\item \x{gawk}: a GNU implementation of \x{awk} with some extensions
\item \x{pdftotext}: probably the most reliable open-source utility in Linux
  for extracting text from PDF files
\item \x{iconv}: a utility for converting text between different encoding
  formats
\item \x{aspell}: an interactive spell checker supporting different languages
  and file formats
\end{enumerate}

\begin{figure}[H]
  \centering
  \includegraphics[scale=0.95]{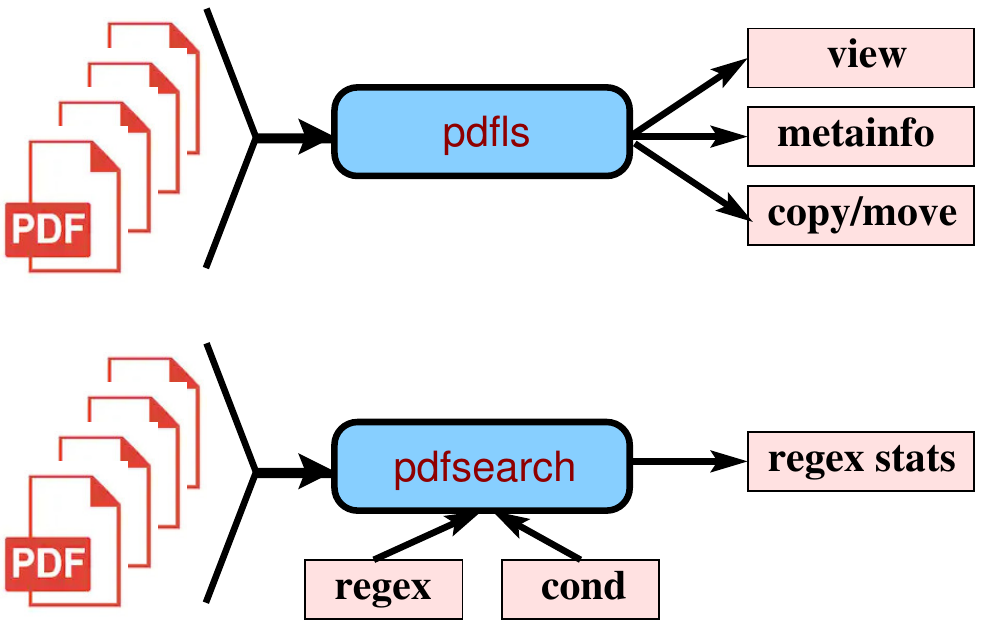}
  \caption{The basic functionality of \x{pdfls} and \x{pdfsearch} shell
    scripts.}\label{pict3}
\end{figure}

\begin{figure}[H]
  \centering
  \includegraphics[scale=0.95]{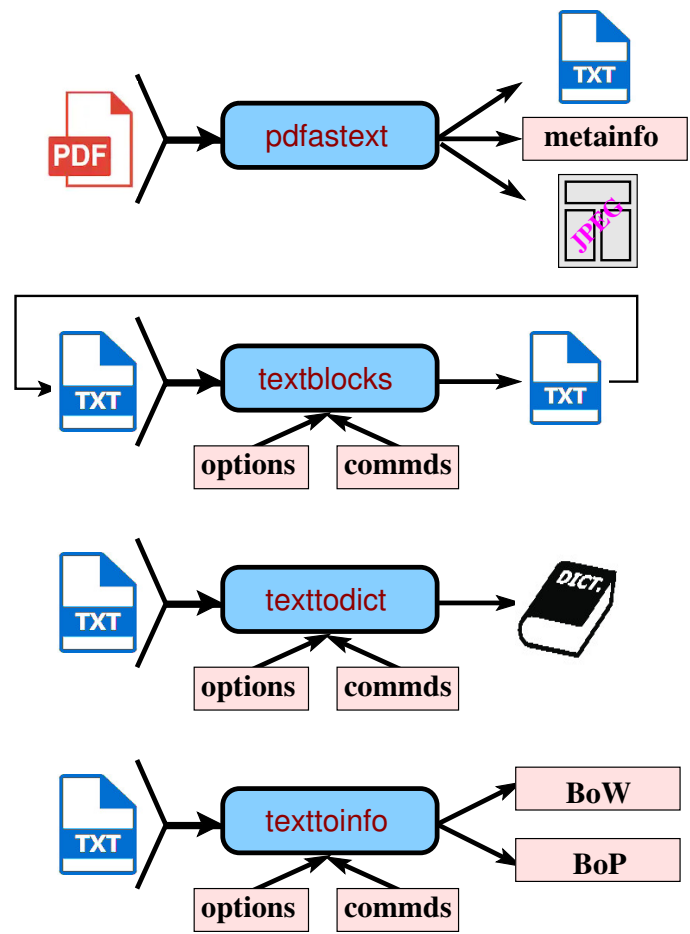}
  \caption{The basic functionality of \x{pdfastext}, \x{textblocks}, and
    \x{texttodict} and \x{texttoinfo} shell scripts.}\label{pict4}
\end{figure}

The \pp\ program consists of 6 shell scripts offering different complimentary
functions. The basic functionality of \x{pdfls} and \x{pdfsearch} utilities is
sketched in \fref{pict3}. These two scripts are typically used for batch
processing and viewing of multiple PDF files. The other 4 shell scripts, i.e.,
\x{pdfastext}, \x{textblocks}, \x{texttoinfo} and \x{texttodict} are intended
to process a single input file. The basic functionality of these 4 other
scripts is shown in \fref{pict4}. All scripts can be invoked to display more
detailed usage instructions. The latest version of \pp\ software is freely
available for download and testing from the Github public
repository\footnote{\url{https://github.com/ploskot/pdfPapers}}. The content of
the repository is briefly described in the appendix. The examples from the
repository are described in the next section.

\header{\x{pdfls}} is a shell script for viewing collections of PDF files which
is a very common task when exploring scientific papers. Instead of opening all
PDF files at once, \x{pdfls} automatically opens the next PDF file once the
previous one was closed. The opening order respects the order of files given as
the input argument. This enables to open the files in a desired order.
\x{pdfls} also supports the interactive mode where a single key press inside
the shell can cause the currently opened file to be copied, moved or skipped.

\header{\x{pdfsearch}} allows to search for complex regular patterns across
collections of PDF files, and then show the overall statistics of the frequency
of occurrence of these regular expression patterns. The actual search is
performed on the text extracted from every PDF file given as the input argument
using the default \x{pdftotext} utility. The PDF to text extraction is done
automatically. The patterns for regular expressions can go beyond what is
implicitly provided by the BASH shell. In particular, the regular expression
sub-patterns can be combined into logical expressions using logical \x{AND} and
\x{OR} operators and comparison operators to test the required or expected
number of occurrences of each sub-pattern within the text. The expressions can
be arbitrarily nested using parentheses. The command output provides
information on how many times the logical pattern was satisfied for each input
file given. There are several options to properly format the generated output.
Multiple PDF files can be queried at once. The shell script implementation
appears to be very fast.

\header{\x{pdfastext}} is a shell script wrapper for the standard \x{pdftotext}
utility. It tries to remedy some deficiencies of \x{pdftotext} and also add
some new features. In particular, the text file generated by \x{pdftotext} is
curated for non-printable characters which can be deleted or transcribed, and
the words and sentences split across lines can be merged together. An
additional file containing meta-information such as the number of pages in the
input PDF file, the author and the producer of the PDF file, the number of
words and characters on each page, and the location of bounding boxes can be
produced. By default, the extracted text is composed of logical units referred
to as blocks. The blocks reflect how the text contents were laid out on the
page by the PDF creation software. Consequently, the blocks of text can differ
vastly from the natural logical flow of the textual contents as desired by a
human reader. The blocks are labeled as the decimal numbers, $N.M$, where $N$
is the page number and $M$ is the block counter within the page. Furthermore,
in order to visualize the block labels and their locations on the page,
\x{pdfastext} can also generate a graphical image of every page with the text
blocks overlayed on the original PDF page in the background.

\header{\x{textblocks}} is a shell script for manipulating blocks of text which
were produced by \x{pdfastext}. The changes can be done on the input file, or a
new file can be produced. Due to the complexity of processing, this script also
provides extensive logging of all operations carried out and other informative
messages into a log file or to the standard terminal output. This is useful for
debugging and to understand unexpected outcomes of the processing. The script
\x{textblocks} can provide information on the blocks contained in the input
file, and check if the blocks are complete (i.e., having both opening and
closing tags and an assigned unique label) and sorted by their label. A
sophisticated block addressing scheme utilizing ranges allows to perform the
operations on given combinations of blocks, pages or on the whole file. The
non-printable characters can be transliterated, or the characters in selected
blocks can be changed to lower-case or upper-case. Any character can be
replaced or appended with a specified string, and the selected strings can be
replaced with other strings. For example, it is possible to break the text in
selected blocks into words or sentences, replace multiple spaces with a single
space, delete leading or trailing spaces from lines, and delete empty lines.
Another option can produce statistics for the selected blocks about their
number of words, the number of words not in a spelling dictionary, and the
number of non-printable characters. The \x{textblocks} script can insert new
blocks at a given location (e.g., before or after the existing block). The new
blocks can be empty or contain a given string. The blocks can be copied or
moved to a new location within the input file, or to the output file. The block
labels can be changed, or orderly renumbered. The blocks can be sorted by their
label. The selected blocks can be deleted, or merged into one of the existing
blocks. Finally, the text inside selected blocks can be filtered with a given
function or another shell script.

\header{\x{texttoinfo}} is a shell script to perform text mining tasks. In the
current version of \pp, only the BoW with the frequency of occurrence and the
multi-word phrase extraction of a given length surrounding the keyword defined
by a regular expression have been implemented. It is recommended that a clean
text file is passed as the input to this script. Since the text mining tasks
are usually the most time consuming, in future versions of \pp, it may be
better to implement text mining algorithms in other programming languages which
are faster such as Java, Python or C/C++.

\header{\x{texttodict}} is a shell script for creating dictionaries for
\x{aspell} or for creating simple lists of words. The list of words can be
obtained from the input text file as a BoW or as a dump of the existing
\x{aspell} dictionary.

\header{Typical workflow} starts from exploring the contents of collected PDF
files using the \x{pdfsl} and \x{pdfsearch} utilities. Both utilities are
straightforward to use. Their command-line calls are intuitive, and their
implementation is fast. They enable to narrow down the focus on a relatively
small number of PDF documents which may be worth exploring more deeply. The
information contents of the selected PDF documents should be explored one by
one. In the first step, the PDF file is converted to a text file using the
\x{pdfastext} utility. Many text blocks on the page often contain supporting
information which can be safely discarded prior to text mining. It is
recommended to first copy the relevant blocks into a new text file using the
\x{textblocks} utility. The page previews showing text block layouts for every
page can be obtained in the first step with \x{pdfastext} utility. The text can
be further cleaned as required using \x{textblocks}. In the last step, the
frequency of occurrence statistics of multi-word phrases are obtained by
running the \x{texttoinfo} command.

\section{Results}

The \pp\ shell script utilities were used to extract the most frequently
occurring phrases or 1 to 4 words in 5 selected papers in biology and life
sciences. The keyword extraction from the papers is presented in subsections
4.1 to 4.5. In addition, the list of biological terms was created by extracting
all words and phrases from the index and a table of acronyms in a PDF e-book.
It is described in subsection 4.6. The shell scripts as well as data files for
all examples considered can be found in the public Github repository (cf.
Appendix).

\subsection{Example 1}

The extraction of the most frequent multi-word phrases was performed for the
paper \cite{sample1}. It is a relatively short paper consisting of 6 pages. The
paper contains mathematical symbols and equations, 5 figures, but no tables. In
addition to the standard content elements such as the paper and section titles,
authors names and affiliations, the paper contains 4 author suggested keywords
(``synthetic circuits, optimal filtering, noise cancellation, adaptive
design''), the statement about the author contributions, acknowledgment, and a
box summarizing the paper significance. A summary of the process of generating
the multi-word phrases of 1 to 4 words is given in \tref{tb:11}. The whole
process was completed in 62s.

\begin{table}[H]
  \caption{Generation of multi-word phrases for paper
    \cite{sample1}}\label{tb:11}
  \centering
  \begin{tabular}{ccccl}
    \hline Level & Phrases & Count & Run time & Output file  \\ \hline
    1 & single word & 200 & 1s & \x{ex5-sample1.w1} \\
    2 & 2-words & 200 & 1s & \x{ex5-sample1.w2} \\
    3 & 3-words & 1863 & 25s & \x{ex5-sample1.w3} \\
    4 & 4-words & 2703 & 35s & \x{ex5-sample1.w4} \\ \hline
  \end{tabular}
\end{table}

\begin{table}[H]
  \caption{The multi-word phrases and their counts identified in paper
    \cite{sample1}}\label{tb:12} 
  \centering\tablesize{\footnotesize}\setlength{\tabcolsep}{3pt}
  \begin{tabular}{|ccl|cl|cl|cl|}\hline
    \multicolumn{3}{|c|}{Level 1} & \multicolumn{2}{c|}{Level 2} &
    \multicolumn{2}{c|}{Level 3} & \multicolumn{2}{c|}{Level 4}\\\hline
67	 & 1.817	 & filter	& 26  & poisson filter	& 10  & ensemble poisson filter	& 6  & provided in si appendix \\
37	 & 2.760	 & circuit	& 20  & death process	 & 6  & signal of interest	& 6  & described in si appendix \\
36	 & 1.424	 & noise	& 16  & optimal filter	 & 6  & number of plasmid	& 4  & vitro using dna strand \\
35	 & 2.518	 & appendix	& 16  & birth rate	 & 6  & dna strand displacement	& 4  & using dna strand displacement \\
33	 & 1.804	 & rate	& 14  & system identification	 & 6  & constitutive promoter pmc	& 4  & tolerate a substantial degree \\
29	 & 1.471	 & sensor	& 14  & noise cancellation	 & 4  & vitro using dna	& 4  & substantial degree of model \\
24	 & 1.636	 & section	& 13  & sensor reaction	 & 4  & using dna strand	& 4  & signal of interest z \\
23	 & 0.977	 & time	& 13  & differential equation	 & 4  & strand displacement cascades	& 4  & shown in si appendix \\
22	 & 2.246	 & birth	& 12  & optimal filters	 & 4  & stochastic simulations of	& 4  & sensor rate c y \\
22	 & 0.563	 & through	& 12  & optimal filtering	 & 4  & sensor time points	& 4  & number of plasmid copies \\
21	 & 1.587	 & estimator	& 10  & synthetic circuits	 & 4  & sensor rate c	& 4  & modeled as a birth \\
21	 & 0.930	 & filtering	& 10  & strand displacement	 & 4  & remarkably high precision	& 4  & in vitro using dna \\
20	 & 1.712	 & filters	& 10  & kalman filter	 & 4  & number of plasmids	& 4  & found in si appendix \\
19	 & 2.426	 & optimal	& 10  & in vitro	 & 4  & information about z	& 4  & dna strand displacement cascades \\
19	 & 1.086	 & process	& 10  & ensemble poisson	 & 4  & inducible promoter pmi	& 4  & degree of model mismatch \\
18	 & 2.560	 & ensemble	& 10  & cell cycle	 & 4  & birth and death	& 4  & death process z 2 \\
18	 & 1.226	 & biochemical	 & 8  & transcription rate	 & 4  & affected by contextual	& 4  & circuit in escherichia coli \\
18	 & 1.128	 & dynamics	 & 8  & time points	 & 4  & adaptive system identification	& 4  & birth and death rates \\
17	 & 4.191	 & circuits	 & 8  & optogenetic circuit	 & 4  & able to estimate	& 4  & attached to a constitutive \\
17	 & 1.169	 & signal	 & 8  & mmse estimator	 & 3  & used an optogenetic	& 4  & approach in vitro using \\\hline
  \end{tabular}
\end{table}

\tref{tb:12} shows the multi-word phrases of 1 to 4 keywords and their
frequency of occurrence which were identified in paper \cite{sample1}. The
number of phrases shown in \tref{tb:12} is 20 for each level. The total number
of phrases generated at each level is shown in \tref{tb:11}. Note that all
words in \tref{tb:12} have been converted to lower-case letters. For the first
level, the second column gives the normalized spreads of given single word
keywords within the paper which were calculated using eq. \eref{eq:1}. Note
that there is a striking difference between having only 4 keywords which were
provided by the authors, and having over 80 phrases across 4 levels given in
\tref{tb:12} to describe and understand the paper. The authors provided
keywords are likely sufficient for reliably indexing the paper in the paper
databases. However, in order to understand the scientific and information
contents of the paper clearly requires to consider many more phrases which are
not restricted by their length.

The scripts and the input and output files used in this example can be obtained
from the sub-directory \x{example01/} located in the \pp\ Github repository
(cf. Appendix).

\subsection{Example 2}

The extraction of the most frequent multi-word phrases was performed for the
paper \cite{sample2}. This is a longer paper having 11 pages, but only 3
displayed mathematical equations and 2 displayed chemical reaction equations.
The paper structure is otherwise standard with several statements given at the
end of the paper just before the references. There are only 3 figures with
captions, and no tables. The authors specify 7 key phrases (``Mathematical
model, Predictive model, Fundamental physical laws, Phenomenology,
Membrane-bounded compartment, T-cell receptor, Somitogenesis clock''). A
summary of the process of generating the multi-word phrases of 1 to 4 words is
given in \tref{tb:21}. The whole process was completed in 84s.

\begin{table}[H]
  \caption{Generation of multi-word phrases for paper
    \cite{sample2}}\label{tb:21}
  \centering
  \begin{tabular}{ccccl}
    \hline Level & Phrases & Count & Run time & Output file  \\ \hline
    1 & single word & 200 & 1s & \x{ex5-sample2.w1} \\
    2 & 2-words & 200 & 2s & \x{ex5-sample2.w2} \\
    3 & 3-words & 2336 & 34s & \x{ex5-sample2.w3} \\
    4 & 4-words & 3300 & 47s & \x{ex5-sample2.w4} \\ \hline
  \end{tabular}
\end{table}

\begin{table}[H]
  \caption{The multi-word phrases and their counts identified in paper
    \cite{sample2}}\label{tb:22} 
  \centering\tablesize{\footnotesize}\setlength{\tabcolsep}{3pt}
  \begin{tabular}{|ccl|cl|cl|cl|}\hline
    \multicolumn{3}{|c|}{Level 1} & \multicolumn{2}{c|}{Level 2} &
    \multicolumn{2}{c|}{Level 3} & \multicolumn{2}{c|}{Level 4}\\\hline
85	 & 3.165	 & model	& 19  & mathematical model	& 12  & model is correct	& 6  & sensitive factor attachment protein \\
34	 & 2.905	 & biology	& 16  & systems biology	 & 8  & her1  and her7	 & 6  & forward and reverse modeling \\
32	 & 3.407	 & models	& 14  & negative feedback	 & 8  & fundamental physical laws	& 6  & descriptions of our pathetic \\
30	 & 4.700	 & assumptions	& 14  & mass action	 & 8  & based on fundamental	& 5  & factor attachment protein receptor \\
26	 & 3.530	 & cell	& 12  & reverse modeling	 & 6  & heinrich and rapoport	& 4  & specific protein tyrosine kinase \\
22	 & 1.491	 & mathematical	& 12  & cell receptor	 & 6  & forward and reverse	& 4  & period of 30  minutes \\
22	 & 1.035	 & molecular	& 10  & somitogenesis clock	 & 6  & factor attachment protein	& 4  & objective descriptions of reality \\
19	 & 2.182	 & protein	& 10  & identical compartments	 & 5  & attachment protein receptor	& 4  & negative and positive feedback \\
17	 & 4.074	 & modeling	 & 8  & time delays	 & 4  & protein tyrosine kinase	& 4  & models based on fundamental \\
17	 & 3.100	 & feedback	 & 8  & rapoport model	 & 4  & physics or even	& 4  & model is a logical \\
16	 & 5.581	 & conclusions	 & 8  & physical laws	 & 4  & negative feedback loop	& 4  & law of mass action \\
15	 & 4.217	 & physics	 & 8  & molecular biology	 & 4  & molecular dynamics models	& 4  & latter from the former \\
15	 & 2.422	 & figure	 & 8  & mathematical models	 & 4  & models in biology	& 4  & guarantee that a model \\
14	 & 6.837	 & clock	 & 8  & lewis model	 & 4  & kinetic proofreading scheme	& 4  & fit what you want \\
14	 & 2.111	 & time	 & 8  & fundamental physical	 & 4  & guarantee of logical	& 4  & bind better to coat \\
14	 & 2.039	 & data	 & 7  & feedback loop	 & 4  & descriptions of reality	& 4  & based on fundamental physical \\
13	 & 4.922	 & snares	 & 6  & time delay	 & 4  & better to coat	& 4  & asking whether we believe \\
13	 & 4.666	 & compartments	 & 6  & somite formation	 & 4  & believe its conclusions	& 3  & forward and reverse model \\
13	 & 4.283	 & negative	 & 6  & sensitive factor	 & 4  & believe its assumptions	& 2  & zebrafish with the help \\
13	 & 0.457	 & biological	 & 6  & reverse model	 & 3  & sensitive factor attachment	& 2  & worked well to account \\\hline
  \end{tabular}    
\end{table}

\tref{tb:22} shows the multi-word phrases of 1 to 4 keywords and their
frequency of occurrence which were identified in paper \cite{sample2}. The
number of phrases shown in \tref{tb:22} is 20 for each level. The total number
of phrases generated at each level is shown in \tref{tb:21}. Note that all
words in \tref{tb:22} have been converted to lower-case letters. For the first
level, the second column gives the normalized spreads of given single word
keywords within the paper which were calculated using eq. \eref{eq:1}. As for
the previous example, the number of generated key phrases is significantly
larger than the number of authors nominated keywords.

The scripts and the input and output files used in this example can be obtained
from the sub-directory \x{example02/} located in the \pp\ Github repository
(cf. Appendix).

\subsection{Example 3}

The extraction of the most frequent multi-word phrases was performed for the
paper \cite{sample3}. This paper is different from all other example papers
considered in that it was published more than 30 years ago. Then, sometime
later, the paper was made available as a PDF document. Although a visual
presentation of the paper is appealing, the extraction of text from the PDF
file turned out to be problematic. In addition to many special characters in
non-standard fonts, the text blocks occasionally mix text lines from
neighbouring columns indicating that \x{pdftotext} had failed to recognize the
locations and properly group some lines of the text. Consequently, the text
file generated by \x{pdfastext} had to be checked and several blocks of text
manually corrected. This suggests that the process of making the older
scientific papers available as PDF documents could be improved, otherwise their
conversion to text is less reliable than for more recently published papers.

The paper \cite{sample3} does not contain any authors specified keywords,
statements and even no references. There are 9 figures, some of them very
large, and no tables. A summary of the process of generating the multi-word
phrases of 1 to 4 words is given in \tref{tb:31}. The whole process was
completed in 69s.

\begin{table}[H]
  \caption{Generation of multi-word phrases for paper
    \cite{sample3}}\label{tb:31}
  \centering
  \begin{tabular}{ccccl}
    \hline Level & Phrases & Count & Run time & Output file  \\ \hline
    1 & single word & 200 & 1s & \x{ex5-sample3.w1} \\
    2 & 2-words & 200 & 2s & \x{ex5-sample3.w2} \\
    3 & 3-words & 1980 & 25s & \x{ex5-sample3.w3} \\
    4 & 4-words & 3060 & 41s & \x{ex5-sample3.w4} \\ \hline
  \end{tabular}
\end{table}

\begin{table}[H]
  \caption{The multi-word phrases and their counts identified in paper
    \cite{sample3}}\label{tb:32} 
  \centering\tablesize{\footnotesize}\setlength{\tabcolsep}{3pt}
  \begin{tabular}{|ccl|cl|cl|cl|}\hline
    \multicolumn{3}{|c|}{Level 1} & \multicolumn{2}{c|}{Level 2} &
    \multicolumn{2}{c|}{Level 3} & \multicolumn{2}{c|}{Level 4}\\\hline
53	 & 1.562	 & energy	& 40  & turing machine	& 28  & amount of energy	& 14  & minimum amount of energy \\
40	 & 4.307	 & machine	& 21  & logic gate	& 12  & random thermal motion	& 10  & absence of a ball \\
40	 & 3.515	 & ball	& 19  & fredkin gate	 & 8  & left or right	 & 6  & segment to the left \\
33	 & 2.738	 & computer	& 17  & ball computer	 & 6  & segment of pipe	 & 6  & presence of a ball \\
31	 & 6.548	 & head	& 16  & minimum amount	 & 6  & order to perform	 & 6  & in order to perform \\
31	 & 1.555	 & computation	& 14  & thermal motion	 & 6  & movement of bits	 & 6  & expend as little energy \\
28	 & 3.095	 & information	& 14  & logic gates	 & 6  & held in place	 & 6  & energy as we wish \\
26	 & 1.467	 & state	& 12  & random thermal	 & 6  & frictionless billiard balls	 & 6  & bit onto the tape \\
25	 & 3.898	 & forward	& 12  & billiard balls	 & 6  & expended in order	 & 6  & ball in a particular \\
24	 & 4.930	 & input	& 11  & head molecule	 & 6  & expend as little	 & 6  & at a logic gate \\
23	 & 5.110	 & turing	& 10  & transition rules	 & 6  & enzymatic turing machine	 & 6  & as little energy as \\
22	 & 2.832	 & gate	& 10  & master camshaft	 & 6  & clockwork turing machine	 & 4  & two balls arrive simultaneously \\
21	 & 4.034	 & balls	& 10  & fredkin gates	 & 6  & bits of information	 & 4  & together with tommaso toffoli \\
21	 & 3.929	 & segment	& 10  & billiard ball	 & 5  & expenditure of energy	 & 4  & taking a long time \\
21	 & 3.482	 & logic	 & 8  & uncertainty principle	 & 4  & without any friction	 & 4  & split segment of pipe \\
20	 & 8.331	 & base	 & 8  & static friction	 & 4  & uncertainty principle does	 & 4  & sometimes the enzyme takes \\
20	 & 0.981	 & amount	 & 8  & small amount	 & 4  & two balls arrive	 & 4  & small as we wish \\
19	 & 7.191	 & molecule	 & 8  & reversible turing	 & 4  & state to state	 & 4  & small amount of energy \\
19	 & 4.075	 & tape	 & 8  & input lines	 & 4  & right or left	 & 4  & set of transition rules \\
19	 & 2.656	 & motion	 & 8  & driving force	 & 4  & represent the output	 & 4  & rna strand and releases \\\hline
  \end{tabular}
\end{table}

\tref{tb:32} shows the multi-word phrases of 1 to 4 keywords and their
frequency of occurrence which were identified in paper \cite{sample3}. The
number of phrases shown in \tref{tb:32} is 20 for each level. The total number
of phrases generated at each level is shown in \tref{tb:31}. Note that all
words in \tref{tb:32} have been converted to lower-case letters. For the first
level, the second column gives the normalized spreads of given single word
keywords within the paper which were calculated using eq. \eref{eq:1}.

The scripts and the input and output files used in this example can be obtained
from the sub-directory \x{example03/} located in the \pp\ Github repository
(cf. Appendix).

\subsection{Example 4}

The extraction of the most frequent multi-word phrases was performed for the
paper \cite{sample4}. The paper contains 4 figures and 1 large table.
Fortunately, the text automatically detected in the table was properly
extracted into separate blocks, so they can be excluded from the main text or
merged into one large block for the subsequent text mining. There are no author
defined keywords and no other statements. The references are included at the
end of the paper. A summary of the process of generating the multi-word phrases
of 1 to 4 words is given in \tref{tb:41}. The whole process was completed in
92s.

\begin{table}[H]
  \caption{Generation of multi-word phrases for paper
    \cite{sample4}}\label{tb:41}
  \centering
  \begin{tabular}{ccccl}
    \hline Level & Phrases & Count & Run time & Output file  \\ \hline
    1 & single word & 200 & 1s & \x{ex5-sample4.w1} \\
    2 & 2-words & 200 & 2s & \x{ex5-sample4.w2} \\
    3 & 3-words & 2423 & 36s & \x{ex5-sample4.w3} \\
    4 & 4-words & 3079 & 53s & \x{ex5-sample4.w4} \\ \hline
  \end{tabular}
\end{table}

\begin{table}[H]
  \caption{The multi-word phrases and their counts identified in paper
    \cite{sample4}}\label{tb:42} 
  \centering\tablesize{\footnotesize}\setlength{\tabcolsep}{3pt}
  \begin{tabular}{|ccl|cl|cl|cl|}\hline
    \multicolumn{3}{|c|}{Level 1} & \multicolumn{2}{c|}{Level 2} &
    \multicolumn{2}{c|}{Level 3} & \multicolumn{2}{c|}{Level 4}\\\hline
85	 & 1.692	 & limits	& 55  & integrated circuit	& 12  & limits to computation	& 4  & two or three dimensions \\
39	 & 1.115	 & integrated	& 42  & integrated circuits	& 10  & integrated circuit design	& 4  & transfer between carriers device \\
36	 & 2.843	 & power	& 22  & emerging technologies	 & 6  & ten years ago	& 4  & size and delay variation \\
33	 & 1.613	 & circuit	& 20  & fundamental limits	 & 6  & speed of light	& 4  & scale to large sizes \\
32	 & 2.881	 & energy	& 19  & time limits	 & 6  & modern integrated circuits	& 4  & permission from gold standard \\
31	 & 1.738	 & circuits	& 14  & power consumption	 & 6  & limits to computing	& 4  & nonphysical limits to computing \\
27	 & 3.242	 & quantum	& 12  & technology node	 & 6  & improvements in computer	& 4  & limits on fundamental limits \\
25	 & 1.023	 & technologies	& 12  & quantum computers	 & 5  & modern integrated circuit	& 4  & information transfer between carriers \\
24	 & 3.636	 & design	& 12  & gate dielectric	 & 4  & voltage scaling 56	 & 4  & image redrawn from figure \\
24	 & 1.914	 & computing	& 12  & engineering obstacles	 & 4  & universality circuit delay	& 4  & fundamental limits to computation \\
24	 & 1.903	 & computation	& 10  & supply voltage	 & 4  & transfer between carriers	& 4  & faster than the best \\
24	 & 1.180	 & scaling	& 10  & moore's law	 & 4  & size and delay	& 4  & carriers device gate dielectric \\
21	 & 4.418	 & computers	& 10  & logic gates	 & 4  & semiconductor integrated circuits	& 4  & between carriers device gate \\
19	 & 1.020	 & time	& 10  & fundamental limit	 & 4  & scale to large	& 2  & years and 600  years \\
18	 & 1.644	 & technology	& 10  & circuit design	 & 4  & redrawn from figure	& 2  & works around engineering obstacles \\
17	 & 0.801	 & performance	 & 9  & time limit	 & 4  & reasonably tight limits	& 2  & wires in several square \\
16	 & 1.319	 & interconnect	 & 8  & universal computers	 & 4  & reasonably tight limit	& 2  & wires get slower relative \\
15	 & 3.654	 & transistors	 & 8  & sequential algorithm	 & 4  & quantum information processing	& 2  & wire stacks from 1997 \\
15	 & 2.396	 & manufacturing	 & 8  & power density	 & 4  & permission from gold	& 2  & wider gate dielectric layer \\
14	 & 2.630	 & emerging	 & 8  & physical space	 & 4  & parallel and sequential	& 2  & wider dielectric layers 26 \\\hline
  \end{tabular}
\end{table}

\tref{tb:42} shows the multi-word phrases of 1 to 4 keywords and their
frequency of occurrence which were identified in paper \cite{sample4}. The
number of phrases shown in \tref{tb:42} is 20 for each level. The total number
of phrases generated at each level is shown in \tref{tb:41}. Note that all
words in \tref{tb:42} have been converted to lower-case letters. For the first
level, the second column gives the normalized spreads of given single word
keywords within the paper which were calculated using eq. \eref{eq:1}.

The scripts and the input and output files used in this example can be obtained
from the sub-directory \x{example04/} located in the \pp\ Github repository
(cf. Appendix).

\subsection{Example 5}

The extraction of the most frequent multi-word phrases was performed for the
paper \cite{sample5}. This is a long paper with 21 pages and relatively
complicated structure. There are 16 figures and 2 tables. There are several
statements including the author contributions, acknowledgment, the support
information summaries, and an inset with the authors summary. There are several
displayed mathematical and chemical equations, and some inline mathematical
symbols and expressions. However, the conversion of PDF to a text file was
straightforward, since only a small number of text blocks containing meaningful
information had to be collected for the subsequent text mining. A summary of
the process of generating the multi-word phrases of 1 to 4 words is given in
\tref{tb:51}. The whole process was completed in 244s, more than double in
comparison to all previous papers considered. Likewise, the number of candidate
3-word and 4-word phrases is nearly doubled in comparison with the previous
papers.

\begin{table}[H]
  \caption{Generation of multi-word phrases for paper
    \cite{sample5}}\label{tb:51}
  \centering
  \begin{tabular}{ccccl}
    \hline Level & Phrases & Count & Run time & Output file  \\ \hline
    1 & single word & 200 & 1s & \x{ex5-sample5.w1} \\
    2 & 2-words & 200 & 3s & \x{ex5-sample5.w2} \\
    3 & 3-words & 4019 & 85s & \x{ex5-sample5.w3} \\
    4 & 4-words & 6946 & 155s & \x{ex5-sample5.w4} \\ \hline
  \end{tabular}
\end{table}

\begin{table}[H]
  \caption{The multi-word phrases and their counts identified in paper
    \cite{sample5}}\label{tb:52} 
  \centering\tablesize{\footnotesize}\setlength{\tabcolsep}{3pt}  
  \begin{tabular}{|ccl|cl|cl|cl|}\hline
    \multicolumn{3}{|c|}{Level 1} & \multicolumn{2}{c|}{Level 2} &
    \multicolumn{2}{c|}{Level 3} & \multicolumn{2}{c|}{Level 4}\\\hline
128	 & 2.318	 & inducer	& 105  & inducer concentration	& 16  & transcription and translation	& 14  & size and inducer concentration \\
128	 & 1.494	 & model	 & 58  & gene expression	& 16  & number of lacy	& 12  & burst size and inducer \\
106	 & 2.201	 & cell	 & 44  & rate constants	& 15  & internal inducer concentration	 & 8  & size as a function \\
88	 & 3.140	 & repressor	 & 43  & in vivo	& 14  & size and inducer	 & 8  & pseudo first order rate \\
71	 & 6.637	 & burst	 & 42  & positive feedback	& 14  & active and inactive	 & 8  & models of gene expression \\
69	 & 2.096	 & operator	 & 42  & inducer concentrations	& 12  & stochastic gene expression	 & 8  & mean number of lacy \\
67	 & 2.157	 & state	 & 41  & burst size	& 12  & lattice microbe method	 & 8  & function of inducer concentration \\
59	 & 4.218	 & noise	 & 38  & rate constant	& 12  & inducible genetic switch	 & 8  & between burst size and \\
59	 & 3.104	 & cells	 & 33  & induced state	& 12  & in vivo crowding	 & 8  & between active and inactive \\
59	 & 1.834	 & simulations	 & 32  & state model	& 12  & fully induced state	 & 8  & active and inactive states \\
56	 & 2.399	 & concentration	 & 28  & stochastic simulations	& 11  & shown in figure	 & 6  & uninduced to the induced \\
55	 & 1.693	 & distributions	 & 28  & population distributions	& 10  & transcription burst size	 & 6  & relationship between burst size \\
53	 & 2.182	 & expression	 & 28  & cell cycle	& 10  & pseudo first order	 & 6  & range of inducer concentrations \\
53	 & 1.997	 & rate	 & 24  & operator complex	& 10  & in vivo environment	 & 6  & proteins produced per burst \\
52	 & 1.817	 & using	 & 24  & free operator	 & 8  & range of inducer	 & 6  & number of proteins produced \\
51	 & 2.730	 & lacy	 & 24  & burst frequency	 & 8  & produced per burst	 & 6  & models of stochastic gene \\
51	 & 1.854	 & figure	 & 22  & protein lifetime	 & 8  & probability of rebinding	 & 6  & model of gene expression \\
50	 & 2.779	 & mean	 & 22  & inducer molecules	 & 8  & probability density function	 & 6  & mean number of proteins \\
49	 & 2.412	 & gene	 & 22  & genetic switch	 & 8  & models of gene	 & 6  & inducible lac genetic switch \\
49	 & 2.382	 & protein	 & 21  & coli cell	 & 8  & mean protein lifetime	 & 6  & frequency of transcriptional bursts \\\hline
  \end{tabular}
\end{table}

\tref{tb:52} shows the multi-word phrases of 1 to 4 keywords and their
frequency of occurrence which were identified in paper \cite{sample5}. The
number of phrases shown in \tref{tb:52} is 20 for each level. The total number
of phrases generated at each level is shown in \tref{tb:51}. Note that all
words in \tref{tb:52} have been converted to lower-case letters. For the first
level, the second column gives the normalized spreads of given single word
keywords within the paper which were calculated using eq. \eref{eq:1}.

The scripts and the input and output files used in this example can be obtained
from the sub-directory \x{example05/} located in the \pp\ Github repository
(cf. Appendix).

\subsection{Example 6}

In the last example, a table of acronyms and the index from a biochemistry
e-book in the PDF format were used to create the lists of biochemical phrases.
Such list can be utilized for keyword extraction and other text mining tasks
assuming biochemical literate. The PDF to text conversion was achieved using
the \x{pdfastext} utility. A small number of irrelevant text blocks was then
deleted, for example, those containing page numbers. The non-printable
characters were deleted, and multiple space and empty lines were also removed.
There are 93 one-word, 124 two-word, 79 three-word, 14 four-word, 9 five word
and only 3 six-word acronyms among 322 acronyms in total. This acronym
distribution approximately reflects the distributions of keywords identified in
the previous five examples. Note also that only the acronym definitions were
considered whereas the actual acronyms were removed from the output file.
Utilizing the shell scripts was particularly useful for processing the 53 page
index file, and creating the list of over 11,000 biochemical terms. The scripts
and the input and output files used in this example can be again obtained from
the sub-directory \x{example10/} located in the \pp\ Github repository (cf.
Appendix).

\section{Discussion}

Most publishers require that the authors add a certain minimum number of
keywords to their paper. These keywords are a subjective choice as they reflect
how the authors would like their paper to be perceived and indexed. For
indexing, the keywords with broader coverage are preferred. However, such
general coverage is unsatisfactory when trying to understand information
contents of papers. Since the paper title and abstract are usually made
available even in paid-for journal repositories, it is sometime recommended to
select the keywords covering the rest of the paper and which are not contained
in the paper title and abstract. This improves the efficiency of keywords use,
and subsequently also the paper visibility in searches. The coverage efficiency
can be also improved if there is little information overlap among the keywords.
The keywords could be assigned the level or a category of importance. For
example, having the primary and secondary keywords can enable more robust
information processing applications beyond information retrieval and indexing.

In our experiments, we observed that having about 10's of phrases consisting of
up to 4 words and having the largest frequency of occurrence gives a reasonably
good idea about the unique focus of the paper. The number of 2-word phrases
considered can be as large as 30 or 40, since there usually exist many such
plausible phrases in most papers. The shorter phrases of 1 or 2 words provide
more general view on the paper whereas longer phrases of more than 3 words give
increasingly specialized view on the paper. Even if only 20 single word
keywords with the largest frequency of occurrence can be considered to a be a
sufficiently good description of the paper key topics, it is important to
assume at least 200 such single word keywords to construct the most frequently
occurring 2-word phrases. Instead of simply enumerating the important phrases
as was done in the previous section, the word clouds and word clusters and
other visualization methods may be preferred in some applications.

The following strategies were ignored in our current implementation of the \pp\
utilities, but they could improve the reliability of keyword identification:
\begin{enumerate}
\item The words and their counts can be merged if they have a common prefix.
  The phrases containing words with common prefix can be clustered.
\item The location of candidate keywords or phrases within the paper appears to
  be important. For example, since the keywords tend to occur in the paper
  title, there are often many candidate keywords detected in the list of
  references which can easily bias their frequency counts.
\item Synonymous, similar and otherwise related words can be identified and
  treated as a group instead of individual words.
\item There should be specific rules for acronyms to be counted as keywords.
\item The text left over from displayed mathematical equations and figure
  labels can be detected and removed, since it rarely contains any meaningful
  information. 
\item Merging of split words, sentences and paragraphs could be further
  improved to be more reliable in most situations encountered.
\item It would be useful to specify which features to use for detecting
  keywords and key phrases as one of the script parameters. Then different
  search strategies can be checked for any paper considered, and the human
  observer decides which feature is the most suitable in a given context.
\end{enumerate}
Our experiments showed that the keyword extraction process is surprisingly
robust to the selection of scoring metric. Despite imperfections in PDF to text
conversions, and using a simple frequency of occurrence metric, satisfactory
sets of multi-word phrases can be identified. Using more complex strategies and
their more reliable implementation appear to change the ordering of phrases
locally rather than globally. Consequently, it is important to generate a
sufficient number of phrases using any scoring function to reliably obtain the
most important phrases among the first, say, 20 having the highest scores.

\begin{figure}[H]
  \centering
  \includegraphics[scale=0.7]{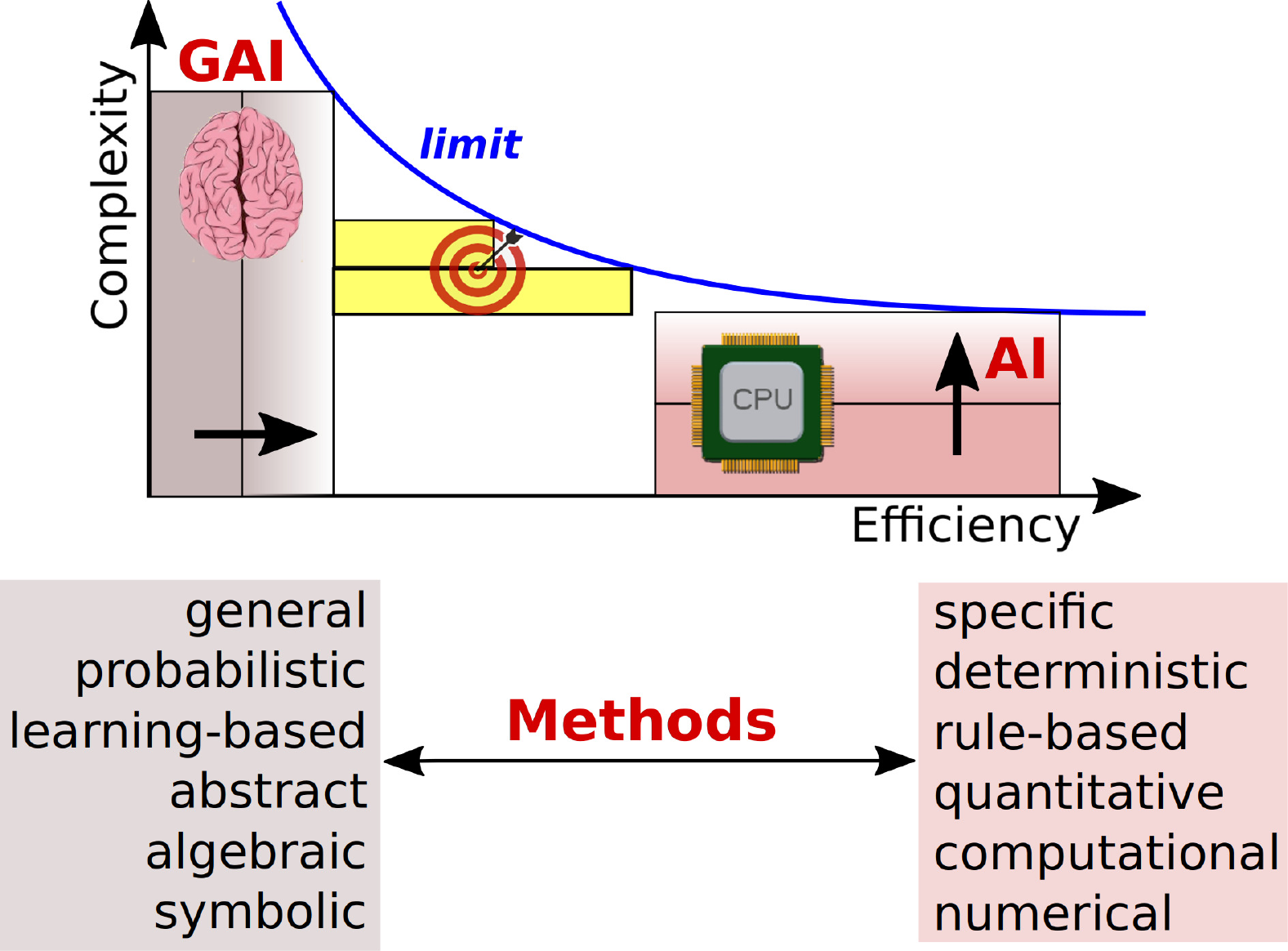}
  \caption{The complexity-efficiency trade-off in problem
    solving.}\label{pict1}
\end{figure}

Having a fully automated system for the keywords identification would be
certainly very desirable. However, the complexity of implementing such system
can grow rapidly, and it may never reach the level of experience of a human
researcher in evaluating scientific papers. The human brain seems to be
extremely good in solving complex problems where the efficiency is not an
issue. On the other hand, the human brain is very inefficient in solving
simpler tasks, particularly if they are of large scale; this is where the
automated systems can serve the researchers very well. As indicated in
\fref{pict1}, our target is the yellow area where we can combine efficient
automated systems with the capability of the human brain. In this paper, the
semi-automated system generates the lists of 1 to 4 word phrases which can be
quickly evaluated by the human researcher to correct for deficiencies in the
algorithm, and to decide which phrases can be discarded despite having large
scores.

The text mining algorithms are time consuming which calls for more efficient
implementation in other languages such as Java, Python or C/C++. Nevertheless,
the core functionality of searching, viewing and extracting text from PDF files
can be provided as shell scripts as long as it is reliable. The ultimate goal
is to develop standard tools for working with PDF documents on *nix systems
which can be accepted to the program repositories for these operating systems.

\section{Conclusions}

The paper reported implementation of shell script utilities to extract the most
frequently occurring multi-word phrases from PDF documents. The utilities are
available for download from the public Github repository named \pp. The core
functionality provided by \pp\ includes a sequential viewing of collections of
PDF files, producing the frequency of occurrence of combined logical and
regular expressions for a group of PDF files, and performing more reliable
conversions of PDF to text. The identification of the most frequent multi-word
phrases was chosen as an example of text mining application. The development of
\pp\ was motivated by the availability of many tools for working with the
contents of text files on *nix (Unix/Linux) systems whereas such tools are very
scarce for PDF documents. Our literature survey showed that there were a lot of
efforts to develop software for working with the contents of PDF files over the
past decade. Many of these programs are either offered online as web
applications, and not as commandline utilities to run locally, or they do not
support group processing of multiple PDF files.

Our strategy for extracting the multi-word phrases is to find longer phrases by
prepending and appending candidate words to the sufficient number of the most
frequently occurring shorter phrases. The word stemming is not consider, and
the word search is made to be case insensitive. Unlike the methods described in
the literature, our experiments suggest that stop-words should only be removed
if they are the first or the last word in a phrase. The automatically
identified frequent phrases can be manually pruned to remove the phrases which
are likely to have small relevance in a given context. This is akin to
combining the power of human brain to solve complex tasks with the efficiency
of computer algorithms to address computing problems at scale.

The developed utilities were demonstrated on finding the most frequently
occurring phrases of 1 to 4 words in 5 selected papers in biology and life
sciences. The text extraction from more recently published papers appear to be
significantly easier than from the older papers. Another example case
demonstrated how to convert the list of acronyms and the index from a PDF
e-book into the list of biological terms which can be used to aid keyword
identification and other text mining tasks. The scripts and data files for all
examples are available in the \pp\ Github public repository.

\vspace{6pt} 



\appendixtitles{no}
\appendix
\section*{Appendix}

The \pp\ repository on Github contains the following directories and files.

\begin{table}[H]
  \centering
  \begin{tabular}{ll}
    \x{example01/} & directory with scripts and outputs for processing the
                     paper \cite{sample1} \\
    \x{example02/} & directory with scripts and outputs for processing the
                     paper \cite{sample2} \\
    \x{example03/} & directory with scripts and outputs for processing the
                     paper \cite{sample3} \\
    \x{example04/} & directory with scripts and outputs for processing the
                     paper \cite{sample4} \\
    \x{example05/} & directory with scripts and outputs for processing the
                     paper \cite{sample5} \\
    \x{example10/} & directory with scripts to create the list of biological
                     terms \\ & from an e-book index \\
    \x{en-dat} & a dump of aspell US-English dictionary \\
    \x{en-stopwords} & a list of common stop words \\
    \x{pdfastext} & a shell script \\
    \x{pdfls} & a shell script \\
    \x{pdfsearch} & a shell script \\
    \x{README.md} & a readme file \\
    \x{textblocks} & a shell script \\
    \x{textblocks.1} & a help file for the shell script \\
    \x{textblocks.t} & examples of function calls for the shell script \\
    \x{texttodict} & a shell script \\
    \x{texttodict.1} & a help file for the shell script \\
    \x{texttoinfo} & a shell script \\
    \x{texttoinfo.1} & a help file for the shell script \\
  \end{tabular}
\end{table}



\reftitle{References}
\externalbibliography{yes}
\bibliography{refer}

\end{document}